\documentclass[usenatbib]{mnras}
\usepackage[T1]{fontenc}
\usepackage{ae,aecompl}

\usepackage[dvips]{graphicx}
\usepackage{graphicx}
\usepackage{amsmath,amssymb}
\usepackage{epsf}
\usepackage{dcolumn}
\usepackage{color}
\usepackage{epsf}
\usepackage{psfrag}
\usepackage{epsfig}
\usepackage{bm}
\usepackage{txfonts}

\title[Probing isotropy with galaxy clusters]{Probing cosmological isotropy with Planck Sunyaev-Zeldovich galaxy clusters}

\author[C. A. P. Bengaly Jr. et al.]{
C. A. P. Bengaly Jr.,$^{1}$\thanks{E-mail: carlosap@on.br}
A. Bernui,$^{1}$\thanks{E-mail: bernui@on.br}
I. S. Ferreira$^{2}$\thanks{E-mail: ivan@fis.unb.br}
J. S. Alcaniz,$^{1}$\thanks{E-mail: alcaniz@on.br}
\\
$^{1}$Observat\'orio Nacional, 20921-400, Rio de Janeiro - RJ, Brasil,\\
$^{2}$Instituto de F\'isica, Universidade de Bras\'ilia, 70910-900, Bras\'ilia - DF, Brasil,\\
}

\date{\today}

\begin{document}
\label{firstpage}
\pagerange{\pageref{firstpage}--\pageref{lastpage}}
\maketitle

\begin{abstract}  
We probe the statistical isotropy hypothesis of the large-scale structure with the second Planck Sunyaev-Zeldovich (PSZ2) galaxy clusters data set. 
Our analysis adopts a statistical-geometrical method which compares the 2-point angular correlation function of objects in antipodal patches of the sky. 
Given possible observational biases, such as the presence of anisotropic sky cuts and the non-uniform exposure of Planck's instrumentation, ensembles of Monte Carlo realisations are produced in order to assess the significance of our results. { When these observational effects are properly taken into account, we find neither evidence for preferred directions in the sky nor signs of large-angle features in the galaxy clusters celestial distribution. The PSZ2 data set is, therefore, in good concordance with the fundamental hypothesis of large-angle isotropy of cosmic objects.}
\end{abstract}


\section{Introduction} \label{intro}

The Cosmological Principle (CP) consists on a fundamental assumption in modern cosmology in which the Universe presents neither special directions, nor special positions, as discussed in~\cite{goodman95, maartens11}, for instance. In this sense, the success of FLRW-based models (especially $\Lambda$CDM paradigm) in explaining the angular power spectrum of the Cosmic Microwave Background (CMB) temperature fluctuations~\citep{planck15a}, the evolution and characterisation of the large-scale structure of the Universe (LSS)~\citep{boss15}, in addition to cosmological distances and ages~\citep{alcaniz99, alcaniz03, simon05, stern10, moresco12, suzuki12, betoule14}, solidified the CP not only as a simplified mathematical hypothesis, but also as a valid physical assumption. Hence, it is crucial to test the CP with observational data for the sake of probing one of the underpinning assumptions of cosmology, besides obtaining a correct interpretation of the physical assumptions underlying the cosmic acceleration and structure formation of the Universe.

In the past years, some analyses reported a possible statistical isotropy violation in large scales, such as: i) CMB temperature fluctuations features (e.g., low multipole alignments, hemispherical power asymmetries, lack of angular correlations at large scales, and the presence of a non-gaussian Cold Spot)~\citep{eriksen04, bmrt, abp, akrami14, bop, planck15b, schwarz16}; ii) large velocity flows from analyses of kinematic Sunyaev-Zeldovich effect in GCs~\citep{kashlinsky09, kashlinsky11, atrio15}, although some controversy have been pointed out by~\cite{osborne11},~\cite{planck14c} regarding the validity of these results\footnote{The possibility of anisotropy in the cosmological expansion was also investigated with Type Ia Supernovae luminosity distance data~\citep{schwarz07, antoniou10, cai12, turnbull12, kalus13, feindt13, jimenez15, appleby15, bengaly15, javanmardi15}. No significant result supporting an anisotropic expansion has been detected so far.}; iii) large number counts dipole anisotropy found in high-$z$ radio sky~\citep{blake02, singal11, rubart13, cobos14, tiwari15, tiwari16}. All these results pose a potential challenge for the concordance model of Cosmology, which relies upon the large-scale isotropy assumption.  

In the light of these puzzles, the validity of the CP must be put under scrutiny with other observational data in order to confirm or refute these results, given that further indications of violation in the statistical isotropy hypothesis would reinforce the demand for a complete reformulation of the concordance model of Cosmology. We can achieve it by directly studying the angular distribution of cosmic objects in antipodal patches of the celestial sphere. In this sense, the second release of Planck Sunyaev-Zeldovich (PSZ2) catalogue~\citep{psz15a, psz15b} provides an ideal observational sample for this purpose, since it comprises over 1600 galaxy clusters (GCs) in a wide sky coverage ($f_{sky} \simeq 0.83$), besides covering a deep redshift range ($z \leq 0.8$). Thus, it corresponds to the most complete all-sky ensemble of GCs available at the present moment. Since these objects are excellent tracers of the large-scale matter distribution of the Universe, they should follow the statistical isotropy which the standard scenario of structure formation is based upon. If any strong evidence for excessive correlations or anti-correlations in the GC distribution is detected, then we could state that the isotropy assumption could be invalid unless explained by limited sky coverage or by the systematics of the observational sample. 

Therefore, the goal of this work is to probe the statistical isotropy assumption by means of a hemispherical comparison of the angular distribution of these PSZ2 catalogue. We also produce synthetic Monte Carlo realisations to account for the anisotropic signatures that could exist in this sample due to the removed area around the galactic plane, in addition to the sky exposure of Planck's observational strategy. The paper is organised as  follows: Section 2 is dedicated to the preparation of the observational sample while Section 3 discusses the methodology developed to carry out the isotropy analyses of the GCs distribution. In Section 4 we discuss the results of these analyses and the tests of statistical significance performed. The concluding remarks and perspectives of this work are presented in Section 5.

\section{Data set} \label{data_set_prep}

\begin{figure*}
\includegraphics[width = 8.3cm, height = 5.5cm]{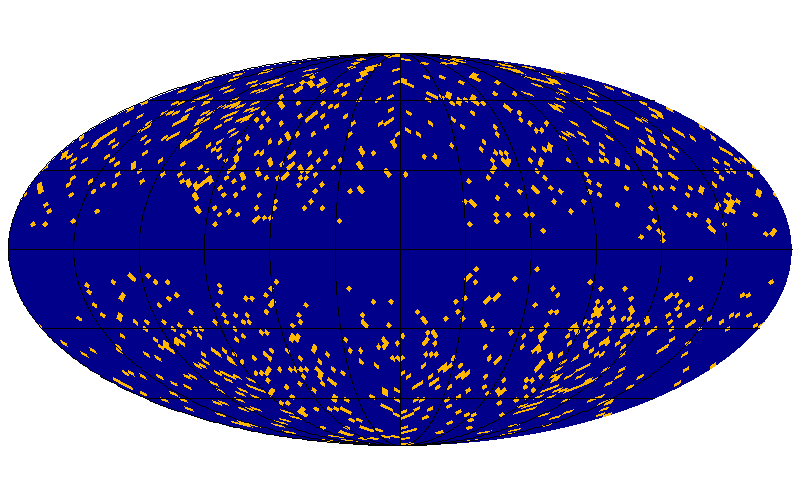}
\includegraphics[width = 8.3cm, height = 5.5cm]{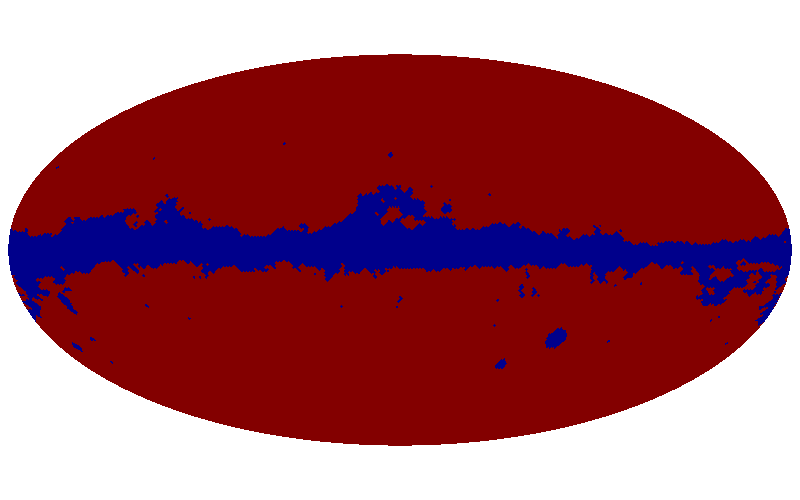}
\vspace{-0.2cm}
\caption{{\it Left panel}: The selected PSZ2 objects for our analyses (see section 2 for details) shown in Mollweide projection. 
{\it Right panel}: The foreground mask of the PSZ2-Union catalogue, whose available area (in red) is $f_{sky} \simeq 0.83$}
\label{fig:map_mask}
\end{figure*}

\begin{figure*}
\includegraphics[width = 8.5cm, height = 7.5cm, angle=180]{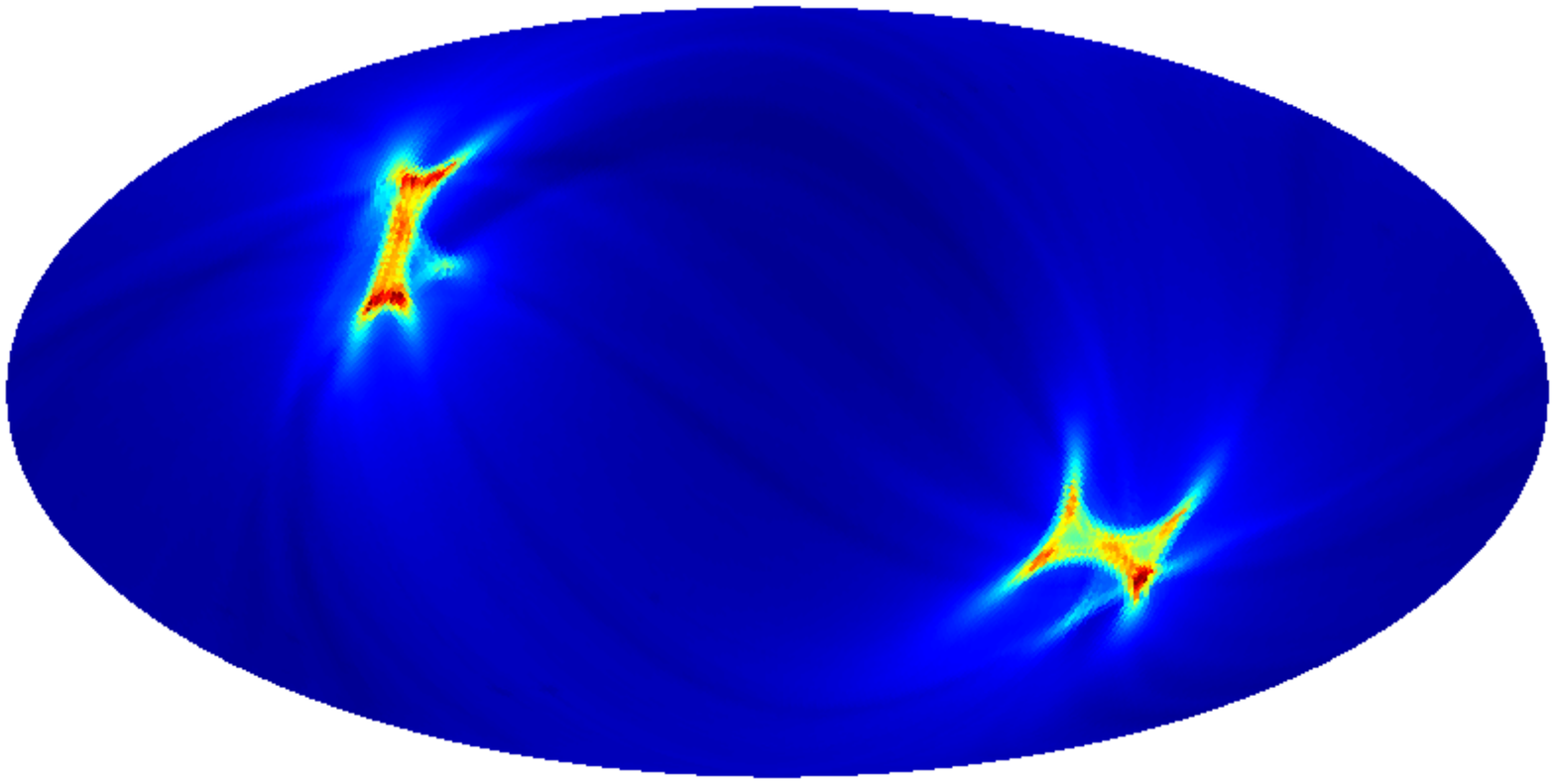}
\includegraphics[width = 8.5cm, height = 7.5cm, angle=180]{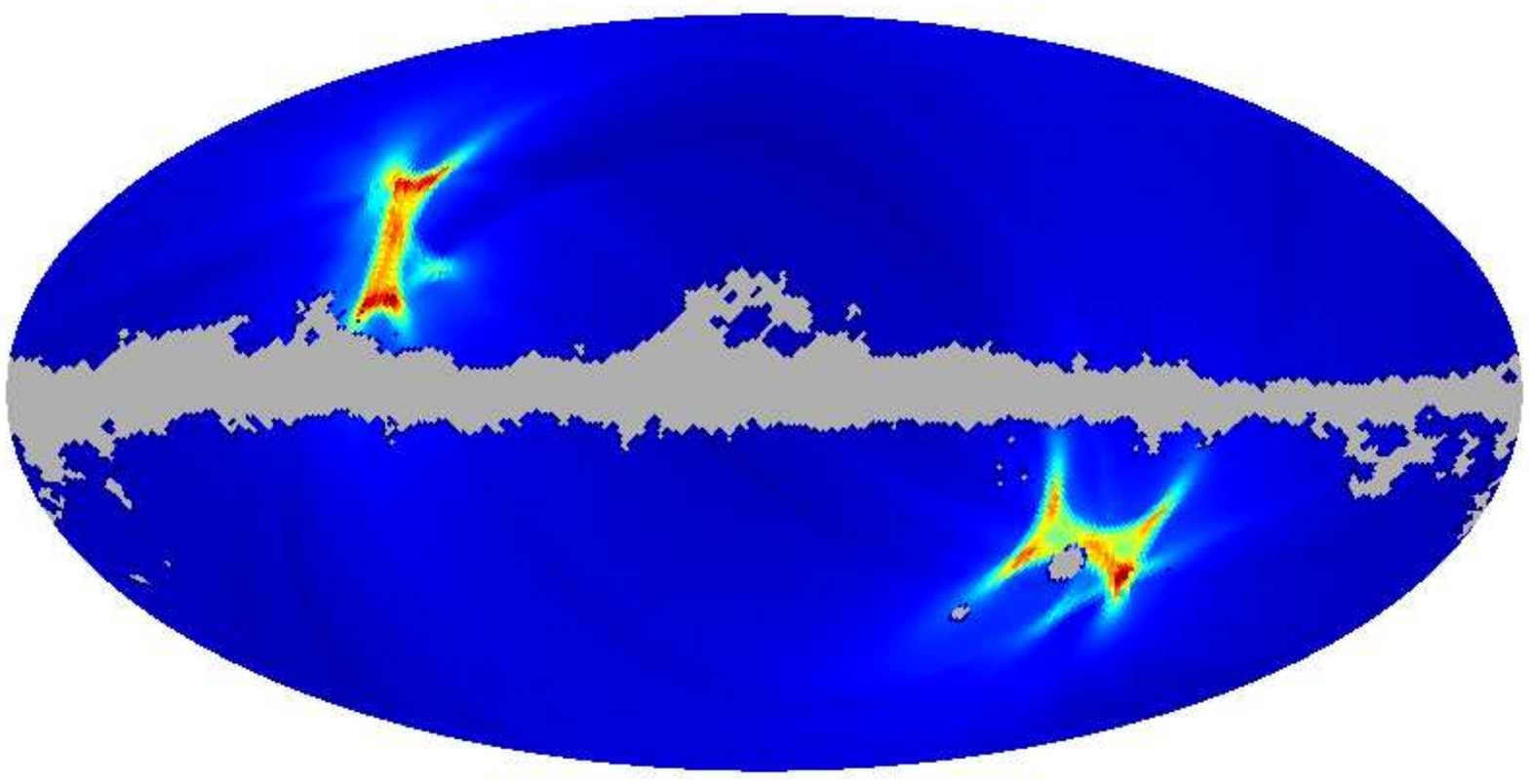}
\vspace{-1.5cm}
\caption{{\it Left panel}: The NUSE map of the Planck satellite, where the brightest spots correspond to the patches of the sky where the Planck's instrument scanned for longer periods. 
{\it Right panel}: The same map after the mask exhibited in Figure~\ref{fig:map_mask} is applied.}
\label{fig:NUSE}
\end{figure*}

Our analyses are performed with the PSZ2 catalogue named {\it Union}, which has been obtained from the IRSA website\footnote{\url{http://irsa.ipac.caltech.edu}}. This set encompasses GCs compiled from the combination of three detection methods based on neural network algorithms (hence the name {\it Union}). The main systematics of this data set consist on spurious infra-red contamination, besides the limitation of neural algorithm in stating whether a GC has been actually detected. Hence, we apply the following queries on the PSZ2 original catalogue:

\begin{itemize}

\item $\mathrm{IR_{FLAG}} < 1$,
\item $\mathrm{Q_{NEURAL}} > 0.4$,
\item $z > -1$.

\end{itemize}

The first two queries eliminate 180 of 1654 objects, which provide a sample purity of 85\%, whereas the last one reduces the sample to 1066 GCs. While the latter is not the main source of uncertainty, we note that the exclusion of the GCs with no redshift determination mildly increases the average signal-to-noise ($S/N$) of their detection, hence corresponding to the most reliable sources of the catalogue\footnote{Note also that there is a sub-set of the PSZ2 catalogue, named {\sc cosmology}, which corresponds to the highest-quality GC detections ($S/N \geq 6$, with a sample purity increased to 95\% in $f_{sky} \simeq 0.65$) data set constructed by Planck team in order to constrain cosmological parameters such as $\Omega_m$ and $\sigma_8$~\citep{psz15b}. We do not focus on this sub-sample since our goal is to test the isotropy assumption with the largest data set possible in terms of the number of GCs and sky coverage.}. The final sample we use throughout this paper is exhibited as yellow dots in the left panel of Figure~\ref{fig:map_mask}, whose corresponding foreground mask leaving an available area of $f_{sky} \simeq 0.83$ appears in the right panel of the same Figure.

\section{Methodology: The Sigma-Map} \label{method}


Here we describe the angular-distribution estimator which leads to quantify deviations from statistical isotropy in a given set of cosmic events with known positions on the celestial sphere~\citep{bmrt,bfw}. Our primary purpose is to illustrate the procedure for defining the discrete function $\sigma$ on the celestial sphere in order to generate its associated map, called {\it sigma-map}. Then, one compares the sigma-map obtained from the data catalogue with a large set of sigma-maps generated from statistically isotropic simulated ensembles in order to assess a measure of (possible) deviation from statistical isotropy in the data set in analysis. 

Let $\Omega_j^{\gamma_0} \equiv \Omega(\theta_j,\phi_j;\gamma_0) \in {\cal S}^2$ be a spherical cap region on the celestial sphere, of $\gamma_0$ degrees of radius, centred at the $j$-th pixel, $j=1, \ldots, N_{\mbox{\tiny caps}}$, where $(\theta_j,\phi_j)$ are the angular coordinates of the center of the $j$-th pixel. Both, the number of spherical caps $N_{\mbox{\tiny caps}}$ and the coordinates of their center $(\theta_j,\phi_j)$ are defined using the HEALPix pixelisation scheme~\citep{gorski05}. The spherical caps are such that their union completely covers the celestial sphere ${\cal S}^2$. We assume galactic coordinates throughout our analyses. Let ${\cal{C}}^{\,j}$ be the catalogue of cosmic objects located in the $j$-th spherical cap $\Omega_j^{\gamma_0}$. The 2-point angular correlation function (2PACF) of these objects~\citep{padmanabhan93}, denoted as $\Delta_j(\gamma_i;\gamma_0)$, is the difference between the normalised frequency distribution and that expected from the number of pairs-of-objects with angular distances in the interval $(\gamma_i - 0.5\delta,\gamma_i + 0.5\delta],\, i=1,\ldots,N_{\mbox{\tiny bins}}$, where $\gamma_i \equiv (i-0.5)\delta$ and $\delta \equiv 2\gamma_0 /N_{\mbox{\tiny bins}}$ is the bin width. The expected distribution is the average of normalized frequency distributions obtained from a large number of simulated realisations of isotropically distributed objects in ${\cal S}^2$, containing the same number of objects as in the data set in analysis. This 2PACF estimator is nothing else than the well known $DD - RR$, termed {\it natural} estimator in the literature~\citep{bt,ab}. A positive (negative) value of $\Delta_j(\gamma_i;\gamma_0)$ indicates that objects with angular separation $\gamma_i$ are correlated (anti-correlated), while zero indicates no correlation. 

Let us define now the scalar function $\sigma: \Omega_j^{\gamma_0} \mapsto {\Re}^{+}$, for $j=1, \ldots, N_{\mbox{\tiny caps}}$, which assigns to the $j$-cap, centred at $(\theta_j,\phi_j)$, a real positive number $\sigma_j \equiv \sigma(\theta_j,\phi_j) \in \Re^+$. We define a measure $\sigma$ of the angular correlations in the $j$-cap as 

\begin{equation} \label{sigma}
\sigma^2_j  \equiv \frac{1}{N_{\mbox{\tiny bins}}}
\sum_{i=1}^{N_{\mbox{\tiny bins}}} \Delta^2_j (\gamma_i;\gamma_0) \, .
\end{equation}

To obtain a quantitative measure of the angular correlation signatures of the GC's sky map, we choose $\gamma_0 = 90^{\circ}$ and cover the celestial sphere with $N_{\mbox{\tiny caps}}=N_{\mbox{\tiny hemis}}=768$ hemispheres, then calculate the set of values 
$\{ \sigma_j, \, j=1,...,N_{\mbox{\tiny caps}} \}$ using eq.~(\ref{sigma}). Patching together the set $\{ \sigma_j \}$ in the celestial sphere according to a coloured scale, we obtain a sigma-map. We quantify the angular correlation signatures of a given sigma-map from a data set by calculating its angular power spectrum. Similar power spectra are calculated, for comparison, with isotropically distributed samples of cosmic objects. 

Since the sigma-map assigns a real number value to each pixel in the celestial sphere, that is, $\sigma = \sigma(\theta,\phi)$, it is possible to expand it in spherical harmonics: $\sigma(\theta,\phi) = \sum_{\ell,\, m} A_{\ell\, m} Y_{\ell\, m}(\theta,\phi)$, where the set of values $\{ \mbox{S}_{\ell} \}$, defined by $\mbox{S}_\ell \equiv (2\ell + 1)^{-1} \sum_{m=-\ell}^{+\ell} |A_{\ell\, m}|^2$, is the angular power spectrum of the sigma-map. Because we are interested in the large-scale angular correlations, we shall concentrate on $\{ \mbox{S}_{\ell}, \,\ell = 1,2,...,10 \}$, i.e., scales larger than $18^{\circ}$ in the sky. Therefore, the small-scale clustering of data points that could arise due to possible visual effects from angular projections, or even non-linearities in the structure formation scenario, is not an issue in our analyses since we are only interested in the large-angle statistics of the sample.

\vspace{-0.3cm}

\section{Statistical significance tests} \label{stat_sig_tests}

\begin{figure*}
\includegraphics[width = 8.3cm, height = 5.5cm]{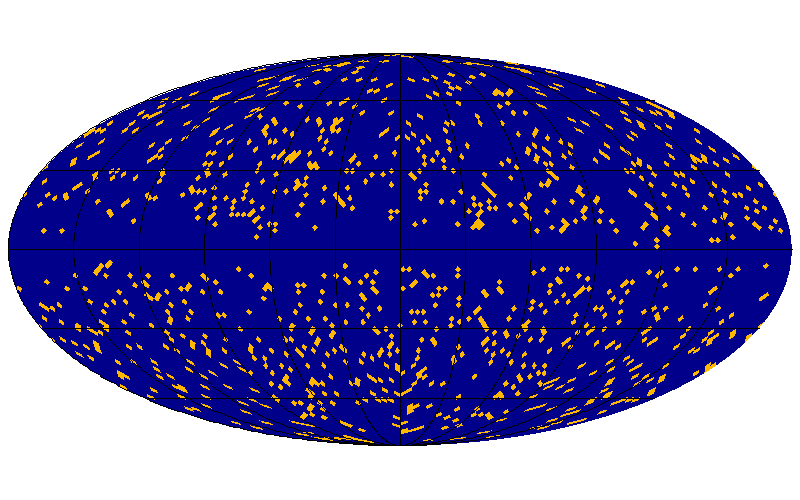}
\includegraphics[width = 8.3cm, height = 5.5cm]{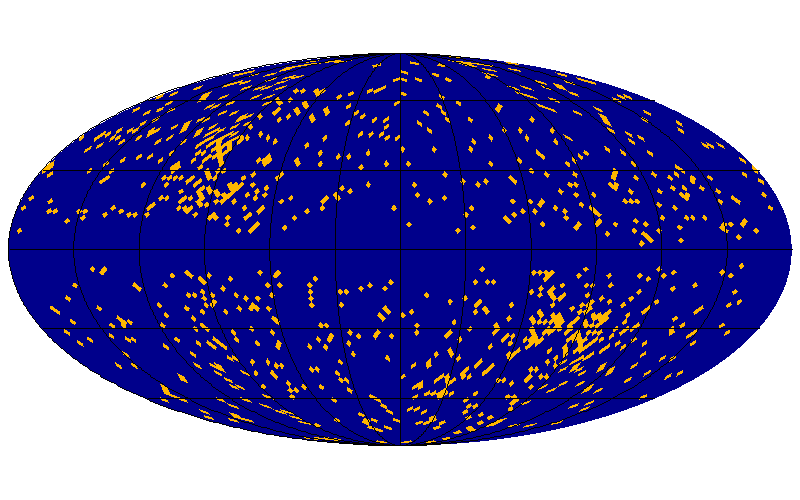}
\vspace{-0.3cm}
\caption{
{\it Left panel}: For illustration, we show the angular distribution of points for a {\it MC iso} realisation map. 
{\it Right panel}: For illustration, we show the angular distribution of a simulated {\it MC aniso} map.
It is possible to note, even by eye, that there are larger concentrations of points toward regions corresponding to the brightest 
spots of the NUSE map displayed in Figure~\ref{fig:NUSE}.}
\label{fig:MC_maps}
\end{figure*}

\subsection{Monte Carlo isotropic simulations}


As previously mentioned, the statistical significance of our analyses is assessed by comparison of the sigma-maps of Monte Carlo (MCs) realisations with the sigma-map computed from the real data. 
We produce an ensemble of 500 statistically isotropic data sets, hereafter termed {\em MCs iso}, with the same available area and approximately the same number of data points as the real data. This is carried out according to the following steps\footnote{As a matter of fact, we sample $n_{\mathrm GCs} \simeq 1284$ in the sky, since this number corresponds to the expected $n_{\mathrm GCs}$ when no foreground mask is considered. In all these cases, the celestial sphere is divided in $49152$ equal area pixels, thus corresponding to a $N_{side}=64$ grid~\citep{gorski05}.}:

\begin{enumerate}

\item Let $\mbox{P}_{\mbox{\tiny I}}$ be a uniform distribution of real random numbers: $\{ p_{\mbox{\tiny I}} \}$, where $p_{\mbox{\tiny I}} \in [0, 1]$ (the sub-index {I} stands for {Isotropy}). Given a specific pixel $j$, we generate a real number $p_{\tiny I}(j) \in \mbox{P}_{\tiny I}$. If $p_{\mbox{\tiny I}} > 1 - (1284 / N_{pix}) \simeq 0.974$, then we assign to this pixel $j$ a GC, otherwise, it remains empty~\footnote{In order to produce $n_{\mathrm GCs} \simeq 1284$ GCs in the full sky each pixel 
should have $\sim \!2.6$\% of probability to be occupied because $(1 - 0.974) \times N_{pix} \simeq 1280$.}. 

\item We repeat such operation for all pixels of the map ($j=1,\cdots,N_{pix}$), thus creating a simulated {\em MCs iso} map with $n_{\mathrm GCs} \simeq 1284$ GCs in the full sky. By repeating this methodology 500 times, we obtain our set of {\em MCs iso} maps. 

\item The final step of this procedure consists in applying the Planck's foreground mask to this set of simulated maps, thereby resulting in a map with $n_{\mathrm GCs} \simeq 1066$. An example of such {\em MC iso} map, produced under these procedures, is displayed in the left panel of Figure~\ref{fig:MC_maps}\footnote{Note that the angular distribution of the GCs in these simulated maps has been extensively tested to confirm their statistical isotropy feature (for details of such tests, see~\cite{bvf})}.  

\end{enumerate}

\subsection{Monte Carlo anisotropic simulations}

In this section we detail the construction of the second set of simulated maps, i.e., an ensemble of 500 realisations which take into account the non-uniform sky exposure (NUSE) function of the Planck satellite, henceforth {\em MCs aniso}. For this purpose, we consider the $N_{\mathrm obs}$ information given by the Planck collaboration\footnote{\url{http://pla.esac.esa.int/pla/\#maps}} about the number of times that each pixel has been visited by the probe\footnote{The NUSE map provides the average number of observations for each sky pixel ($N_{\mathrm obs}$) of the four lowest frequency channels of Planck's instrumentation, namely 100, 143, 217, and 353 GHz, according to $N_{\mathrm obs}/N_{\mathrm obs,max}$, where $N_{\mathrm obs,max}$ is the maximal number of observations for each channel. We did not consider the $N_{\mathrm obs}$ maps for the two highest frequencies (545 and 857 GHz) because they present quite lower $N_{\mathrm obs, max}$ values compared to the lower frequency channels, which could bias this normalisation procedure adopted to obtain the final NUSE map. However, we stress that the NUSE should be regarded as an approximation for possible directional weights adopted in Planck's detection algorithm of GCs, as such procedure is beyond the scope of this work.}. We then produce the map with normalised ${\bar N}_{\mathrm obs}$ data: ${\bar N}_{\mathrm obs}(j) \equiv N_{\mathrm obs}(j) / N^{\mbox{\tiny max}}_{\mathrm obs}$, therefore ${\bar N}_{\mathrm obs}(j) \in [0,1], \mbox{for} j = 1, \cdots, N_{\mathrm pix}$, where $N_{\mathrm pix} = 49152$ is the number of pixels in the celestial sphere. The final result of these procedures are displayed in both panels of Figure~\ref{fig:NUSE}.

In order to produce a {\em MC aniso} ensemble, we establish the following steps: 

\begin{enumerate}


\item Let $\mbox{P}_{\mbox{\tiny I}}$ and $\mbox{Q}_{\mbox{\tiny I}}$ be two uniform distributions of real random numbers in the interval $\in [0,1]$. Given a pixel $j$, we generate two real numbers $p_{\mbox{\tiny I}}(j) \in \mbox{P}_{\mbox{\tiny I}}$ 
and $q_{\mbox{\tiny I}}(j) \in \mbox{Q}_{\mbox{\tiny I}}$ and define the number $p_{\mbox{\tiny A}}$ as: $p_{\mbox{\tiny A}}(j) \equiv q_{\mbox{\tiny I}}(j) \times {\bar N}_{\mathrm obs}(j)$ (the sub-index A stands for anisotropy). If $p_{\mbox{\tiny I}}(j) + p_{\mbox{\tiny A}}(j) > 1.5$  or $p_{\mbox{\tiny I}}(j) \gtrsim 0.974$, we assign a GC to this pixel $j$, otherwise, it remains empty. Note that the first condition imposes higher probabilities to assign a GC toward the ecliptic poles (the most intense patches in the $N_{\mathrm obs}$ map in Figure~\ref{fig:NUSE}), while the second one ensures that, outside such ecliptic regions (thus ${\bar N}_{\mathrm obs}(j) \simeq 0$), the isotropy condition predominates. 

\item Performing this operation for all pixels of the map ($j=1,\cdots,N_{pix}$), we obtain a simulated {\em MC aniso} map with $1284$ GCs in the full sky. We repeat this operation 500 times in order to obtain a set of 500 {\em MCs aniso} maps. 

\item We apply the Planck's foreground mask to this set, resulting in a simulated map with $n_{\mathrm GCs} \simeq 1066$. An example of a simulated {\em MC aniso} map produced according to this procedure is displayed, for the sake of illustration, in the right panel of Figure~\ref{fig:MC_maps}.  

\end{enumerate}

Hence, we are able to test the statistical isotropy of the PSZ2 data, considering that a possible source of clustering arises due to this anisotropic sky exposure toward the ecliptic poles, which could be manifested as strong correlations and anti-correlations in the sigma-map analysis, when compared to the typical fluctuations of isotropic distributions, as shown in~\cite{bfw, ukwatta16}. This is done by comparing the PSZ2 sigma-map power spectrum with both {\em MC-iso} and {\em MC-aniso} ensembles, given that this power spectrum realises the angular features of these data sets. By means of this analysis, we can test whether systematic effects, like the NUSE and the foreground mask, could lead to false anisotropic signals, or whether there is indeed an intrinsic anisotropy in the data.

\section{Data analysis and results} \label{DA_results}
 
\begin{figure*}
\includegraphics[width = 8.5cm, height = 6.5cm]{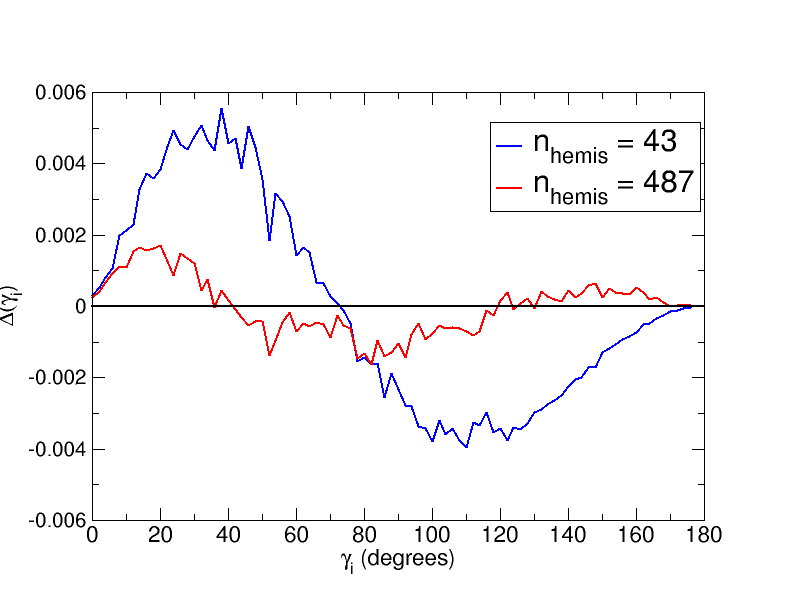}
\includegraphics[width = 8.5cm, height = 6.5cm]{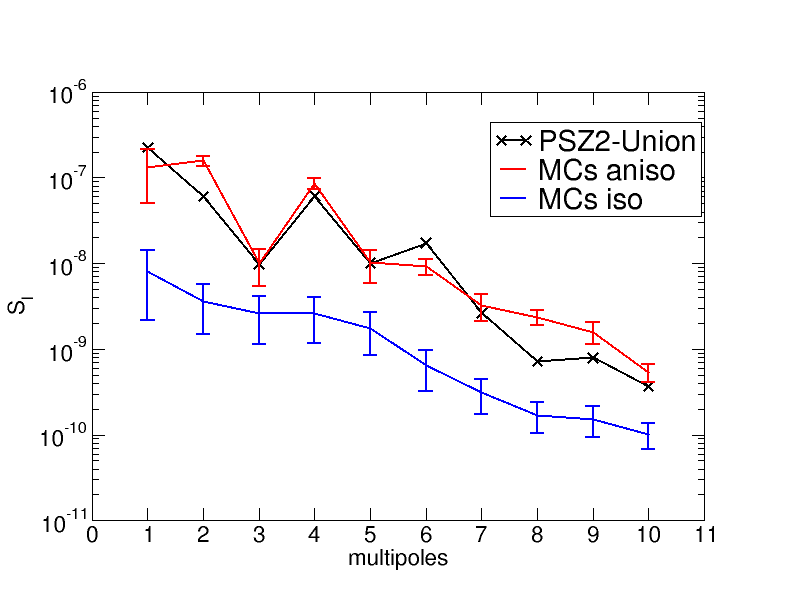}
\caption{
{\it Left panel}: We show the 2PACF $\Delta(\gamma_i)$ for two hemispheres obtained from the PSZ2 catalogue for illustrative reasons. $\mbox{n}_{\mathrm hemis}=43$ denotes the hemisphere position with maximal sigma-map, located at $(l,b) = (45.00^{\circ}, 60.43^{\circ})$, while $\mbox{n}_{\mathrm hemis}=487$ stands for the hemisphere whose minimal sigma-map had been attained, which is centred towards $(l,b) = (247.50^{\circ},-14.48^{\circ})$. {\it Right panel}: The angular power spectrum $\mbox{S}_\ell$ of the PSZ2 sigma-map compared with the average power spectra obtained from the 500 isotropic ({\em MCs iso}, blue curve) and NUSE-based ({\em MCs aniso}, red curve) realisations.}
\label{fig:sigmamap}
\end{figure*}
 
The left panel of Figure~\ref{fig:sigmamap} depicts, for illustrative reasons, the 2PACF $\Delta(\gamma)$ (see Eq.~(\ref{sigma})) obtained in two different hemispheres along the $\gamma$ starting from the hemisphere centre. The hemispheres $\mbox{n}_{\mathrm hemis}=43$ and $\mbox{n}_{\mathrm hemis}=487$, as indicated in this panel as the blue and red curves, respectively, correspond to those where the maximal and minimal sigma values (noting that the sigma denotes the sum of the square of this $\Delta$ over all these $\gamma$) were attained, respectively. 

It is noticeable that the angular correlations fluctuates much more in the former case, hence indicating larger angular correlations (and anti-correlations) in the GCs encompassed in this region. When approximating the sigma-map results obtained through the whole celestial sphere as a dipole, we obtain a direction whose maximal value is located towards the north-western patch of the sky, so, the minimal value points towards the south-east. Even though this direction resembles the maximal CMB power asymmetry localisation, we are unable to ascribe this signal to any of these features in a statistically significant manner. Thus, we provide no support for a significant preferred direction in the PSZ2 map which could be associated with the aforementioned CMB features or the nearby velocity flow. 

Furthermore, we show the power spectrum of the sigma-map, namely $\mbox{S}_{\ell}$, as black crosses in the right panel of Figure~\ref{fig:sigmamap}. The red curve denotes the mean sigma-map power spectra of 500 {\em MCs aniso}, while the blue curve represents the mean spectra from the idealistic isotropic realisations. In all these cases, the central values of these curves are given by their arithmetic average and the error bars correspond to the mean absolute deviation (MAD) of these simulated data sets. We adopt the MAD instead of the standard deviation (STD) because the $\mbox{S}_{\ell}$'s of these 500 MCs are very skewed to the right, i.e., they present a long-tailed distribution, so that the MAD gives a more robust estimation of the uncertainty around their mean value than the STD. We note that the PSZ2 sigma-map behaviour is much closer to the average of 500 {\em MCs aniso} than the isotropic realisations, yet the multipole moments $\mbox{S}_{2}$, $\mbox{S}_{6}$, and $\mbox{S}_{8}$, corresponding to the angular scales $90^{\circ}, 30^{\circ}$, and $22.5^{\circ}$, respectively, are slightly in tension with them. Such mild discrepancies could be ascribed to some other features in the data besides those we are addressing, as possible impurities and contamination that had not been eliminated in our quality tests. 

Nevertheless, we point out that, since these {\em MCs aniso} give the upper limit of the possible amount of anisotropy in each angular scale because of these selection effects, and since the apparent anisotropy of the GCs distribution matches this prediction in most of the angular scales, there is no significant suggestion of anisotropy in the PSZ2 data apart from mild anisotropic features. Accordingly, such signals can be accounted by the asymmetric sky cut and the anisotropic sky exposure. Thus, we can conclude that the GC distribution of the PSZ2 sample is in good concordance with the statistical isotropy hypothesis underlying the standard model of Cosmology.

\section{Conclusions} \label{conc}


In this work, the cosmological isotropy of the large-scale structure of the Universe has been probed with the largest all-sky catalogue of GCs available in the literature, i.e., the second release of Planck Sunyaev Zeldovich sources (PSZ2)~\citep{psz15a, psz15b}. This has been accomplished by mapping the angular distribution of GCs through the celestial sphere using a geometrical-statistical test, named sigma-map. Given the nature of this method, any departure from the statistical isotropy of the data would be revealed by discrepancies of its power spectrum when compared to isotropic data sets. Since we had performed this test in large angular scales, such discrepancy would suggest that the statistical isotropy assumption may not hold, thus the concordance model of Cosmology would need to be completely reformulated.

Our analysis has found no statistically significant indication for anomalous anisotropy in the PSZ2 catalogue, as well as no links with the CMB features or large velocity flows, that have been previously suggested by other authors. As a matter of fact, we have found that the power spectra of the sigma-map computation presents good concordance between the real data and the MC realisations in most of the angular scales probed. Nevertheless, we point out that this agreement happens only if the incomplete sky coverage, and especially the anisotropic sky exposure, are properly incorporated in these simulated data sets. When the latter is neglected, we have noticed that the sigma-map power spectra are substantially lower than the real data's. Since the NUSE provides an upper limit of the anisotropy of the GC sky maps introduced by the Planck's scanning strategy, such observational feature seems to be the main cause of the power excess we have found in PSZ2 large-angle correlations.

We have concluded that the GC angular distribution, which comprises the most massive objects composing the large-scale structure of the Universe, indeed agrees with the cosmic isotropy assumption in large angular scales. Yet, we stress that it is crucial to repeat these tests, or even propose new ones, given the prospect of future GC survey such as eROSITA~\citep{erosita}, which is expected to provide an all-sky sample of $n_{\mathrm GCs} \sim 10^5$, thus enormously enhancing the precision of the test carried out in this work. Furthermore, the advent of much larger galaxy or SN data sets by future surveys, such as LSST~\citep{lsst} and SKA~\citep{ska1,ska2}, may provide assessments of the concordance model and the CP with unprecedented precision in the next decade.


\section*{acknowledgments}
We thank Mariachiara Rossetti for useful discussions about the PSZ2 catalogue. CAPB Jr. acknowledges CAPES for the financial support. AB and ISF acknowledge the {\em Science without Borders Program} of CAPES and CNPq, for a PVE project (88881.064966/2014-01) and a PDE fellowship (234529/2014-08), respectively. JSA acknowledges financial support from CNPq, FAPERJ and INEspa\c{c}o. We also acknowledge the HEALPix package for the derivation of many of the results presented in this work. This research has made use of the NASA / IPAC Infrared Science Archive, which is operated by the Jet Propulsion Laboratory, California Institute of Technology, under contract with the National Aeronautics and Space Administration. The Planck Sunyaev-Zeldovich catalogue used here is based on observations obtained with Planck (\url{http://www.esa.int/Planck}), an ESA science mission with instruments and contributions directly funded by ESA Member States, NASA, and Canada.

\bsp	
\label{lastpage}


\begin{thebibliography}{99}

\bibitem[\protect\citeauthoryear{Abell et al.}{2009}]{lsst}
Abell, P. A. R. {\it et al.} [LSST Red Book 2.0], arXiv:0912.0201
%
\bibitem[\protect\citeauthoryear{Abramo et al.}{2009}]{abp}
Abramo, L. R. {\it et al.}, 2009, JCAP 12, 013
%
\bibitem[\protect\citeauthoryear{Ade et al.}{2014}]{planck14a}
Ade, P. A. R. {\it et al.} [Planck collaboration], Astronom. Astrophys., 2014, 571, A23 
%
\bibitem[\protect\citeauthoryear{Ade et al.}{2014}]{planck14c}
Ade, P. A. R. {\it et al.} [Planck collaboration], Astronom. Astrophys., 2014, 561, A97 
%
\bibitem[\protect\citeauthoryear{Ade et al.}{2015a}]{planck15a} 
Ade, P. A. R. {\it et al.} [Planck collaboration], 2015a, arXiv:1502.01589
%
\bibitem[\protect\citeauthoryear{Ade et al.}{2015b}]{psz15a} 
Ade, P. A. R. {\it et al.} [Planck collaboration], 2015b, arXiv:1502.01597
%
\bibitem[\protect\citeauthoryear{Ade et al.}{2015c}]{psz15b} 
Ade, P. A. R. {\it et al.} [Planck collaboration], 2015c, arXiv:1502.01598
%
\bibitem[\protect\citeauthoryear{Ade et al.}{2015d}]{planck15b} 
Ade, P. A. R. {\it et al.} [Planck collaboration], 2015d, arXiv:1506.07135
%
%
\bibitem[\protect\citeauthoryear{Akrami et al.}{2014}]{akrami14}
Akrami, Y. {\it et al.}, 2014, Astrophys. J., 784, L42
%
\bibitem[\protect\citeauthoryear{Alcaniz \& Lima}{1999}]{alcaniz99}
Alcaniz, J. S. \& Lima, J. A. S., 1999, Astrophys. J., 521, L87
%
\bibitem[\protect\citeauthoryear{Alcaniz et al.}{2003}]{alcaniz03}
Alcaniz, J. S., Lima, J. A. S. \& Cunha, J. V., 2003,  Mon. Not. Roy. Astron. Soc., 340, L39 
%
\bibitem[\protect\citeauthoryear{Antoniou \& Perivolaropoulos}{2010}]{antoniou10}
Antoniou, I. \& Perivolaropoulos, L., 2010, JCAP, 12, 012
%
\bibitem[\protect\citeauthoryear{Appleby et al.}{2015}]{appleby15}
Appleby, S. {\it et al.}, 2015, Astrophys. J., 801, 2, 76 
%
\bibitem[\protect\citeauthoryear{Atrio-Barandela et al.}{2015}]{atrio15}
Atrio-Barandela, F. {\it et al.}, 2015, Astrophys. J., 810, 2
%
\bibitem[\protect\citeauthoryear{Aubourg et al.}{2015}]{boss15} 
Aubourg, E. {\it et al.} [BOSS collaboration], 2015, Phys. Rev. D, 92, 123516
%
\bibitem[\protect\citeauthoryear{Bengaly et al.}{2015}]{bengaly15}
Bengaly, C. A. P.. {\it et al.}, 2015, Astrophys. J., 808, 39
%
\bibitem[\protect\citeauthoryear{Bernui \& Teixeira}{1999}]{bt}
Bernui, A. \& Teixeira, A. F. F., 1999, astro-ph/9904180
%
\bibitem[\protect\citeauthoryear{Bernui et al.}{2004}]{bvf} 
Bernui, A., Villela, T. \& Ferreira, I., 2004, Int. J. Mod. Phys. D, 13, 1189
%
\bibitem[\protect\citeauthoryear{Bernui}{2005}]{ab}
Bernui, A., 2005, Braz. J. Phys., 35, 1185
%
\bibitem[\protect\citeauthoryear{Bernui et al.}{2007}]{bmrt} 
Bernui, A. {\it et al.}, 2007, Astron. Astrophys., 464, 479 
%
\bibitem[\protect\citeauthoryear{Bernui et al.}{2008}]{bfw} 
Bernui, A., Ferreira, I. S. \& Wuensche, C. A., 2008, Astrophys. J., 673, 968
%
\bibitem[\protect\citeauthoryear{Bernui et al.}{2014}]{bop}
Bernui, A. {\it et al.}, 2014, JCAP, 10, 041 
%
\bibitem[\protect\citeauthoryear{Betoule et al.}{2014}]{betoule14} 
Betoule, M. {\it et al.}, 2014, Astron. Astrophys., 568, A22
%
\bibitem[\protect\citeauthoryear{Blake \& Wall}{2002}]{blake02} 
Blake, C. \& Wall, J., 2002, Nature, 416, 150
%
\bibitem[\protect\citeauthoryear{Cai \& Tuo}{2012}]{cai12}
Cai, R. G. \& Tuo, Z., 2012, JCAP, 02, 004
%
\bibitem[\protect\citeauthoryear{Clarkson}{2012}]{clarkson12}
Clarkson, C., 2012, Comp. Rend. de l'Acad des Sci., 13, 682
%
\bibitem[\protect\citeauthoryear{Eriksen et al.}{2004}]{eriksen04}
Eriksen, H. K. {\it et al.}, 2004, Astroph. J., 605, 14
%
\bibitem[\protect\citeauthoryear{Feindt et al.}{2013}]{feindt13}
Feindt, U. {\it et al.}, 2013, Astron. Astrophys. 560, A90
%
\bibitem[\protect\citeauthoryear{Fern\'andez-Cobos et al.}{2014}]{cobos14}
Fern\'andez-Cobos, R. {\it et al.}, 2014, Mon. Not. Roy. Astron. Soc., 441, no.3, 2392
%
\bibitem[\protect\citeauthoryear{Goodman}{1995}]{goodman95}
Goodman, J., 1995, Phys. Rev. D, 52, 1821
%
\bibitem[\protect\citeauthoryear{G\'orski et al.}{2005}]{gorski05}
G\'orski, K. M. {\it et al.}, 2005, Astrophys. J., 622, 759
%
\bibitem[\protect\citeauthoryear{Javanmardi et al.}{2015}]{javanmardi15}
Javanmardi, B. {\it et al.}, 2015, Astrophys. J., 810, 47
%
\bibitem[\protect\citeauthoryear{Jim\'enez et al.}{2015}]{jimenez15}
Jim\'enez, J. N. {\it et al.}, 2015, Phys. Lett. B, 741, 168
%
\bibitem[\protect\citeauthoryear{Kalus et al.}{2013}]{kalus13}
Kalus, N. {\it et al.}, 2013, Astron. Astrophys., 513, A56
%
\bibitem[\protect\citeauthoryear{Kashlinsky et al.}{2009}]{kashlinsky09}
Kashlinsky, A. {\it et al.}, 2009, Astrophys. J., 691, 1479
%
\bibitem[\protect\citeauthoryear{Kashlinsky et al.}{2011}]{kashlinsky11}
Kashlinsky, A. {\it et al.}, 2011, Astrophys. J., 732, 1
%
%
\bibitem[\protect\citeauthoryear{Maartens}{2011}]{maartens11} 
Maartens, R., 2011, Phil. Trans. R. Soc. A, 369, 5115 
%
\bibitem[\protect\citeauthoryear{Maartens et al.}{2015}]{ska1}
Maartens, R. {\it et al.}, arXiv:1501.04076
%
\bibitem[\protect\citeauthoryear{Merloni et al.}{2012}]{erosita}
Merloni, A. {\it et al.} [eROSITA Science Book], 2012, arXiv:1209.3114
%
\bibitem[\protect\citeauthoryear{Moresco et al.}{2012}]{moresco12}
Moresco, M. {\it et al.}, 2012, JCAP, 08, 006
%
\bibitem[\protect\citeauthoryear{Osborne et al.}{2011}]{osborne11}
Osborne, S. {\it et al.}, 2011, Astrophys. J., 737, 98
%
\bibitem[\protect\citeauthoryear{Padmanabhan}{1993}]{padmanabhan93} 
Padmanabhan, T., 1993, Structure formation in the Universe, Cambridge Univ. Press
%
\bibitem[\protect\citeauthoryear{Rubart \& Schwarz}{2013}]{rubart13}
Rubart, M. \& Schwarz, D. J., 2013, Astron. Astrophys., 555, A117
%
\bibitem[\protect\citeauthoryear{Schwarz \& Weinhorst}{2007}]{schwarz07}
Schwarz, D. J. \& Weinhorst, B., 2007, Astron. Astrophys., 474, 717
%
\bibitem[\protect\citeauthoryear{Schwarz et al.}{2016}]{schwarz16}
Schwarz, D. J. {\it et al.}, 2016, Class. Quant. Grav. 33, 18, 184001
%
\bibitem[\protect\citeauthoryear{Schwarz et al.}{2015}]{ska2}
Schwarz, D. J. {\it et al.}, 2015, arXiv:1501.03820 
%
\bibitem[\protect\citeauthoryear{Simon et al.}{2005}]{simon05} 
Simon, J. {\it et al.}, 2005, Phys. Rev. D, 71, 123001 
%
\bibitem[\protect\citeauthoryear{Singal}{2011}]{singal11}
Singal, A., 2011, Astrophys. J., 722, L23
%
\bibitem[\protect\citeauthoryear{Stern et al.}{2010}]{stern10}
Stern, D. {\it et al.}, 2010, JCAP, 02, 008
%
\bibitem[\protect\citeauthoryear{Suzuki et al.}{2012}]{suzuki12}
Suzuki, N. {\it et al.}, 2012, Astroph. J., 746, 85
%
\bibitem[\protect\citeauthoryear{Tiwari et al.}{2015}]{tiwari15}
Tiwari, P. {\it et al.}, 2015, Astropart. Phys., 61, 1
%
\bibitem[\protect\citeauthoryear{Tiwari \& Nusser}{2016}]{tiwari16}
Tiwari, P. \& Nusser, A., 2016, JCAP, 03, 062
%
\bibitem[\protect\citeauthoryear{Turnbull et al.}{2012}]{turnbull12}
Turnbull, S. J. {\it et al.}, 2012, Mon. Not. Roy. Astron. Soc., 420, 447
%
\bibitem[\protect\citeauthoryear{Ukwatta \& W\'ozniak}{2016}]{ukwatta16}
Ukwatta, T. \& W\'ozniak, P., 2016, Mon. Not. Roy. Astron. Soc., 455, 703

\end{thebibliography}
\end{document}